\documentclass[12pt]{article}

\usepackage{latexsym}

\usepackage{graphics}
\usepackage{graphicx}
\usepackage{epstopdf}

\begin{document}


\title{Magnetized Black Hole on Taub-Nut Instanton}

\author{
    Petya G. Nedkova$^{1}$\thanks{E-mail:pnedkova@phys.uni-sofia.bg}, Stoytcho S. Yazadjiev$^{1,2}$\thanks{E-mail: yazad@phys.uni-sofia.bg}\\
{\footnotesize ${}^{1}$ Department of Theoretical Physics,
                Faculty of Physics, Sofia University,}\\
{\footnotesize  5 James Bourchier Boulevard, Sofia~1164, Bulgaria }\\
{\footnotesize ${}^{2}$ Theoretical Astrophysics,
                        Eberhard Karls University of T\"{u}bingen,}\\
{\footnotesize  Auf der Morgenstelle 10, 72076 T\"{u}bingen, Germany } \\}

\date{}

\maketitle

\begin{abstract}
We present an exact solution to the 5D Einstein-Maxwell-dilaton
equations describing a static black hole on Taub-Nut instanton. By
construction the solution does not possess a charge, but is
magnetized along the compact dimension. As a limit we obtain a new
regular solution representing a magnetized Kaluza-Klein monopole. We
investigate the relevant physical properties and derive the
Smarr-like relations.
\end{abstract}

\section{Introduction}

In recent years it has been demonstrated that higher dimensional
gravity admits a variety of solutions with nontrivial geometry.
Among the  different configurations, certain black holes were
constructed, which were described as 'sitting on' gravitational
instantons. This class of solutions includes the so called black
holes on Kaluza-Klein bubbles
\cite{Elvang:2002}-\cite{Nedkova:2010}, as well as black holes on
Taub-Nut, Taub-Bolt, Kerr- and Eguchi-Hanson instantons
\cite{Ishihara:2005}-\cite{Ishihara:2006}. A recent review on the
topic is available written by Chen and Teo \cite{Chen:2011}.

The instanton solutions possess interesting physical properties induced by their complicated geometry.
In a recent work we discussed the thermodynamics of vacuum and electrostatic black holes on asymptotically locally flat gravitational instantons \cite{Yazad:2011}.
The goal of the current paper is to achieve some progress in the investigation of their behavior in magnetic fields.

Magnetized black holes have attracted a lot of attention in
astrophysics since it is considered that they can provide viable
models for realistic stellar-mass and supermassive black holes.
Different mechanisms of electromagnetic energy extraction from
rotating magnetized black hole have been proposed, the
Blandford-Znajek one considered the most relevant
\cite{Blandford:1977}, hoping to explain the formation of the highly
relativistic jets from galactic nuclei. Other interesting physical
phenomena were discovered as well concerning black holes in magnetic
fields \cite{Aliev:1989a}. Such is the gravitational analog of the
Meissner effect which consists in the expulsion of the magnetic flux
lines from black holes horizons as they approach extremality
\cite{Karas:1991}-\cite{Gibbons:1998}, and the charge accretion
leading to the charging up of a rotating black holes immersed in
external magnetic field \cite{Wald:1974}-\cite{Aliev:2004}. The
scattering and Hawking radiation of magnetized black holes were also
actively investigated, as well as the motion of charged particles in
their vicinity \cite{Aliev:2002}-\cite{Frolov:2011a}. It was
demonstrated that the super-radiant instability exhibited by
rotating black holes and the intensity of the Hawking evaporation is
amplified in the presence of magnetic field \cite{Konoplya:2009},
\cite{Kokkotas:2011}. Very recently it was argued that particles
with high center-of-mass energy can be produced as a result of
certain particle collisions in the vicinity of a weakly magnetized
non-rotating black hole \cite{Frolov:2011b}. Thus magnetized
non-rotating black holes could serve as particle
accelerators under some conditions.

Exact solutions to the Einstein-Maxwell equations provide valuable intuition for examining black hole astrophysics.
Magnetized black hole solutions were constructed early in four dimensional spacetime \cite{Ernst:1976}-\cite{Aliev:1989b} by applying  Harrison transformation.
Recently they were generalized to a variety of solutions to the 5D Einstein-Maxwell, and Einstein-Maxwell-dilaton equations describing black objects in external
magnetic fields \cite{Ortaggio:2004}, \cite{Yazad:2005}. Since only the simplest solution representing black hole on gravitational instanton, the black hole
on a Kaluza-Klein bubble, has been magnetized so far \cite{YN1}, we consider that it is important to obtain further magnetized solutions belonging to this class.

The paper is organized as follows. In the first section we present a
new exact solution to the 5D Einstein-Maxwell equations representing
a static magnetized black hole on a Taub-Nut instanton. We examine
its limits and obtain another solution of physical importance - a
magnetized version of the Kaluza-Klein monopole. Next, we
investigate the physical properties of the solution, and calculate
its mass and tension using both Komar integrals and the counter-term
method and comparing the results. The nut charge and potential are
obtained as well, using the relations demonstrated in
\cite{Yazad:2011} and generalizing the definition of the nut
potential appropriately for the current case. Section 4 is devoted
to a rigorous derivation of the relevant Smarr relations.

\section{Exact Solution}
We consider  the Einstein-Maxwell-dilaton gravity (EMd) in
$5$-dimensional spacetime  with the action

\begin{equation}
I = {1\over 16\pi} \int d^5x \sqrt{-g}\left(R - 2g^{\mu\nu}\partial_{\mu}\varphi \partial_{\nu}\varphi  -
e^{-2a\varphi}F^{\mu\nu}F_{\mu\nu} \right),
\end{equation}
which leads to the field equations
\begin{eqnarray} \label{FE}
R_{\mu\nu} &=& 2\partial_{\mu}\varphi \partial_{\nu}\varphi + 2e^{-2a\varphi} \left[F_{\mu\rho}F_{\nu}^{\rho} - {1\over 6}g_{\mu\nu} F_{\beta\rho} F^{\beta\rho}\right], \\
\nabla_{\mu}\nabla^{\mu}\varphi &=& -{a\over 2} e^{-2a\varphi} F_{\nu\rho}F^{\nu\rho}, \nonumber \\
&\nabla_{\mu}&\left[e^{-2a\varphi} F^{\mu\nu} \right]  = 0, \nonumber
\end{eqnarray}
where $R_{\mu\nu}$ is the Ricci tensor for the spacetime metric $g_{\mu\nu}$, $F_{\mu\nu}$ is the Maxwell tensor,
$\varphi$ is the dilaton field and $a$ is the dilaton coupling parameter.

In the present paper we are interested in EMd solutions admitting
three commuting Killing vectors, one asymptotically timelike Killing
vector  $\xi$, and two spacelike Killing vectors $\eta$ and $k$ or
more precisely, solutions with  a group of symmetry  $R\times
U(1)^2$.  We focus on pure magnetic solutions with $i_\xi F=0$ and
nonzero magnetic potentials $\Phi_{\eta}=i_{\eta}F$ and $\Phi_{k}=
i_{k}F$. In this case and for dilaton coupling parameter
$a=\sqrt{8/3}$ we have found the following exact solution to the
field equations
\begin{eqnarray}\label{solution}
ds^2 &=& V^{1\over3}(r)\left[ - (1 - \frac{r_+}{r})dt^2 + \frac{r + r_0}{r - r_+}dr^2 + r(r + r_0)\left(d\theta^2 + \sin^2\theta d\phi^2\right)\right] + \nonumber \\
 &&V^{-{2\over3}}(r)\frac{r}{r + r_0}\left(d\psi + r_\infty \cos\theta d\phi \right)^2, \nonumber \\
 e^{-a\varphi} &=& V^{2\over3}(r), \nonumber \\
\Phi_k &=& -\frac{\lambda}{2}\frac{r}{r + r_0}V^{-1}(r), \nonumber \\
\Phi_\eta &=& \Phi_k r_\infty \cos\theta.
\end{eqnarray}
where metric function $V(r)$ is given by
\begin{eqnarray}
V(r) &=& \frac{1}{1 + \lambda^2}\left(1 + \frac{\lambda^2 r}{r +
r_0}\right).
\end{eqnarray}
Here $-\infty <\lambda < \infty $ , $0\le r_{+}<\infty$, $0\le
r_{0}<\infty $ are parameters and  $r_\infty$ is defined by
\begin{eqnarray}
r_\infty &=& \sqrt{\frac{r_0(r_0 + r_+)}{1 + \lambda^2}}.
\end{eqnarray}

The Maxwell 2-form $F$ is given by

\begin{eqnarray}\label{EF}
F = d\psi\wedge d\Phi_k + d\phi\wedge d\Phi_\eta.
\end{eqnarray}

In the coordinates of the solution the Killing vectors are given by
$\xi=\partial/\partial t$, $\eta=\partial/\partial \phi$ and
$k=\partial/\partial \psi$.

As the expression reveals,  the electromagnetic vector potential is
directed along the 1-form corresponding to the compact dimension,
which is parameterized by the angular coordinate $\psi$.

In the limit $\lambda\to 0$ the magnetic field vanishes and the
solution reduces to the vacuum black hole on a Taub-Nut instanton
\cite{Ishihara:2005}. It is also interesting to consider another
limit by setting $r_+=0$. In this case we obtain a completely
regular metric in the form

\begin{eqnarray}
ds^2 &=& V^{1\over3}\left[ -dt^2 + \frac{r + r_0}{r}dr^2 + r(r + r_0)\left(d\theta^2 + \sin^2\theta d\phi^2\right)\right] + \nonumber \\
 &&V^{-{2\over3}}\frac{r}{r + r_0}\left(d\psi + r_\infty \cos\theta d\phi \right)^2, \nonumber \\
\end{eqnarray}
It represents a magnetized generalization of Kaluza-Klein monopole
discovered by  \cite{Gross:1983}, and Sorkin \cite{Sorkin:1983}.

The solution  possesses a horizon located at $r = r_+$ and its
spacelike cross sections at $r=const.$ are diffeomorphic to a Hopf
fibration of $S^3$. Taking also into account   the natural limits of
the solution mentioned above we can interpret our solution as a
magnetized  black hole on a Taub-Nut instanton.

The interval structure of the solution is the following (see fig.1):

\begin{itemize}
\item a semi-infinite space-like interval located at $\left( r \geq r_+, \theta = \pi \right)$ with direction $l_L = (0, r_{\infty}, 1)$;
\item a finite timelike interval located at $\left( r = r_+, 0 \leq\theta \leq\pi \right)$ with direction $l_H=\frac{1}{\kappa_H}(1,0,0)$ corresponding to the black hole horizon;
\item a semi-infinite space-like interval at $\left( r \geq r_+, \theta = 0 \right)$ with direction $l_R = (0, -r_{\infty}, 1)$.
\end{itemize}
The directions of the intervals are determined by their coordinates with respect to a basis of Killing
vectors $\{\frac{\partial}{\partial t}, \frac{\partial}{\partial \psi}, \frac{\partial}{\partial \phi}\}$.
The length of the $S^1$ fibre at infinity is equal to $L = 4\pi r_\infty$, and $\kappa_H$ is the surface gravity of the horizon.

\begin{figure} [h]
\begin{center}
      \includegraphics[width=12.cm]{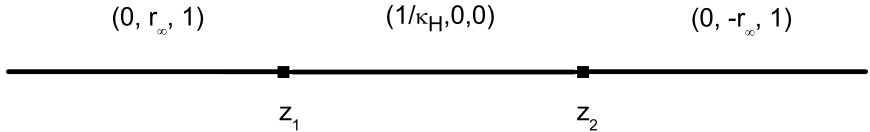}
\caption{Interval structure of magnetized black hole on Taub-Nut instanton}
      \end{center}
\end{figure}
Note that the parameters of the interval structure are not directly
inherited from the vacuum black hole on Taub-Nut instanton since the
parameter $r_\infty$ is modified by the presence of magnetizing
parameter $\lambda$.

In the spirit  of the uniqueness theorem of  \cite{Yazad:2010}, the
solution is completely determined not only by its interval structure
but also by an appropriately  defined magnetic flux. In the case
under consideration we can use the magnetic flux $\Psi$ through the
base space $S^2_{\infty}$ of the $S^1$--fibration at infinity,
namely

\begin{eqnarray}\label{flux}
\Psi=\int_{S^2_{\infty}}F=L\Phi_{k}(\infty)= -2\pi\lambda
r_{\infty}=-2\pi\lambda \sqrt{\frac{r_{0}(r_{0}+ r_{+})}{1+
\lambda^2}}.
\end{eqnarray}

\section{Physical Quantities}

\subsection{Mass and Tension}

The solution is characterized by two conserved gravitational charges
- the mass and the tension \cite{Traschen:2001}, \cite{Traschen:2003}, which can be calculated either by
generalized Komar integrals \cite{Townsend:2001}, \cite{YN1}, or by
the counterterm method \cite{Kraus: 1999}. In the latter approach we
consider the counterterm introduced by Mann and Stelea
\cite{Mann:2005} (see also \cite{Radu:2006})
\begin{equation}\label{counterterm}
I_{ct} = {1\over 8\pi} \int d^4x \sqrt{-h} \sqrt{2{\cal R}},
\end{equation}
leading to a boundary stress-energy tensor in the form
\begin{equation}\label{SET}
T_{ij} = {1\over 8\pi} \left[ K_{ij} - K h_{ij} - \Omega ({\cal
R}_{ij} - {\cal R}h_{ij} ) - h_{ij} D^kD_{k}\Omega +
D_{i}D_{j}\Omega \right],
\end{equation}
where  K is the trace of the extrinsic curvature $K_{ij}$ of the
boundary, ${\cal{R}}$ and $D_k$ are the Ricci scalar and the covariant
derivative with respect to the boundary metric $h_{ij}$, and
$\Omega= \sqrt{2\over {\cal R}}$.

We will use both methods in our computations and show that they lead to equivalent results. The generalized Komar integrals for the mass and the tension are defined as

\begin{eqnarray}\label{MT}
M_{ADM} &=&  - {L\over 16\pi} \int_{S^{2}_{\infty}} \left[2i_k \star d\xi - i_\xi \star d k \right], \\
{\cal{T}} &=&  - {1\over 16\pi} \int_{S^{2}_{\infty}} \left[i_k \star d\xi - 2i_\xi \star d k \right], \nonumber
\end{eqnarray}
where $\xi = \frac{\partial}{\partial t}$ is the Killing field associated with time translations, $k  = \frac{\partial}{\partial\psi}$ is the Killing field corresponding to the compact dimension, $L$ is the length of the $S^1$ fibre and $S^2_{\infty}$ is the base space of $S^1$-fibration at infinity. By direct calculation we obtain the result

\begin{eqnarray}\label{MN}
M_{ADM} &=&  \frac{L}{2}\left(r_{+} + {1\over 2}r_{0}\right)   \\
{\cal{T}} &=& \frac{1}{4} \left(r_{+} + \frac{2 + \lambda^2}{1+\lambda^2}r_{0}\right). \nonumber
\end{eqnarray}

On the other hand, we can calculate the relevant components of the stress-energy tensor

\begin{eqnarray}
8\pi T^{t}_{t} &=& {1\over r^2} \left({1\over 2}r_{0} + r_{+} \right) + {\cal O}({1\over r^3}),   \\
8\pi T^{\psi}_{\psi} &=& {1\over 2r^2} \left(r_{+} + \frac{2 + \lambda^2}{1+\lambda^2}r_{0} \right) + {\cal O}({1\over r^3}). \nonumber
\end{eqnarray}

According to the counterterm method the conserved quantities are obtained from the boundary stress-energy tensor as

\begin{equation}
{\cal Q} = \int_{\Sigma} d\Sigma_{i} T^{i}_{j}\xi^{j},
\end{equation}
where $\xi$ is a Killing vector generating an isometry of the boundary. The conserved quantity represents the mass in the case when $\xi = \partial/\partial t $, and the tension, when $\xi = \partial/\partial \psi$ \cite{Stelea:2009}. Thus, we obtain

\begin{eqnarray}
&&M_{ADM} = \frac{1}{8\pi}\int\left({1\over 2}r_{0} + r_{+} \right)\sin{\theta}d\theta d\phi d\psi ,  \\
&&{\cal{T}} = \frac{1}{16\pi}\int\left(r_{+} + \frac{2 + \lambda^2}{1+\lambda^2}r_{0}\right)\sin{\theta}d\theta d\phi.
\nonumber
\end{eqnarray}
which leads to the same result as (\ref{MN}) after performing the integration. Although the expression for the ADM mass formally coincides with the corresponding one in the vacuum case \cite{Ishihara:2005},\cite{Yazadjiev:2006}, it should be recognized that the value of the parameter $L$ is different, since it is affected by the external magnetic field.

In addition to the ADM mass, an intrinsic mass of the black hole can be introduces by the Komar integral

\begin{eqnarray}\label{MH}
M_H = - {L\over 16\pi} \int_{H} \left[2i_k \star d\xi - i_\xi \star d k \right].
\end{eqnarray}
which in our case obtains the explicit form
\begin{eqnarray}
M_H = \frac{L}{2}r_+.
\end{eqnarray}
The black hole mass can be expressed also in terms of the horizon area $A_H$ and surface gravity $\kappa_H$ as

\begin{eqnarray}\label{MH}
M_H = \frac{1}{4\pi}\kappa_H A_H.
\end{eqnarray}

The surface gravity on the black hole horizon is determined by

\begin{equation}
\kappa_H = \sqrt{-{1\over2}\xi_{\mu;\nu}\xi^{\mu;\nu}}|_H,
\end{equation}
where $\xi = \partial/\partial t $ is the timelike Killing field. It leads to the result

\begin{equation}
\kappa_H = \frac{2\pi}{L}\sqrt{\frac{r_0}{r_+ (1 +
\lambda^2)}}=\frac{1}{2\sqrt{r_{+}(r_{0} + r_{+})}},
\end{equation}
The area of the horizon is calculated as

\begin{eqnarray}
A_H = \int_H \sqrt{g_H}d\theta d\phi d\psi =  L^2r_+\sqrt{\frac{r_+
(1 + \lambda^2)}{r_0}}= \frac{16\pi^2}{\sqrt{1+
\lambda^2}}r^{3/2}_{+}r^{1/2}_{0}(r_{+}+ r_{0}).
\end{eqnarray}
It is obvious by the explicit expressions that (\ref{MH}) is satisfied.

\subsection{Nut charge and potential}

The spacial boundary at infinity of the solution manifold is diffeomorphic to a nontrivial $S^1$ bundle over $S^2$, therefore the solution possesses a nut charge. It is defined by the Komar-like integral \cite{Hunter:1998}

\begin{equation}
N = - {1\over 8\pi} \int_{C^2} d\left(\frac{k}{{\cal V}}\right),
\end{equation}
where $k$ is the Killing 1-form associated with the $S^1$ fibre at
infinity,  ${\cal V}$ is its norm and $C^2$ is a two-dimensional
surface, encompassing the nut. In our case this is equivalent to the
relation

\begin{equation}\label{Nut}
N = {1\over 2}r_\infty = \frac{L}{8\pi},
\end{equation}
which was derived in \cite{Yazad:2011} for black holes on asymptotically locally flat gravitational instantons.

\paragraph{}In addition to the nut charge there exists a related characterstic, called a nut potential.
This is revealed if we examine the 1-form $i_\xi i_k \star dk$, which can be represented in the form \cite{Yazad:2011}

\begin{eqnarray}\label{NUT1}
d i_\xi i_k \star d k &=& 2\star\left[R(k)\wedge k \wedge \xi \right].
\end{eqnarray}

Taking into account that

\begin{eqnarray}
\star R(k) = - 2e^{-2a\varphi} \left( -{2\over 3}i_{k}F\wedge \star F + {1\over 3} F\wedge i_{k}\star F \right),
\end{eqnarray}
and using the explicit form of the electromagnetic field (\ref{EF}), we obtain

\begin{eqnarray}
\star\left[R(k)\wedge k \wedge \xi \right] = 2d\Phi_k\wedge i_\xi
i_k e^{-2a\varphi}\star F .
\end{eqnarray}

It follows from the field equations that $di_\xi i_k
e^{-2a\varphi}\star F = 0$, consequently we can introduce an
electromagnetic potential ${\cal{B}}$ such that $d{\cal{B}} = i_\xi
i_k e^{-2a\varphi}\star F$. Taking advantage of it, eqn.
($\ref{NUT1}$) yields
\begin{eqnarray}
d i_\xi i_k \star d k -  4d\left(\Phi_k d\cal{B}\right)=0 .
\end{eqnarray}

The 1-form $i_\xi i_k \star d k - 4\Phi_k d{\cal{B}}$ is invariant under the Killing fields $ \xi,
k$ and $\eta$ and can be viewed as defined on the factor space $\hat{M}= M/ R \times U(1)^2$.  Since the factor space $ \hat{M}= M/ R
\times U(1)^2$ is simply connected \cite{Hollands:2007}, there exists a globally defined potential $\chi$, such as

\begin{equation}\label{Nut potential}
 d\chi = i_\xi i_k \star d k - 4\Phi_k d{\cal{B}}
\end{equation}

This relation determines the nut potential corresponding to the solution we investigate. It should be noted that
its form distinguishes from the vacuum and electrostatic cases \cite{Yazad:2011} since now it incorporates a term connected with the electromagnetic field.

The nut potential and the electromagnetic potential ${\cal{B}}$ possess the following explicit form

\begin{eqnarray}
\chi &=& \frac{r_\infty(1+\lambda^2)}{r + r_0}, \\ \nonumber
{\cal{B}}&=& \frac{\lambda}{2}\frac{r_\infty}{r + r_0},
\end{eqnarray}
where they are normalized in such a way that they vanish at infinity.

\section{Smarr-like Relations}

In this section we are going to derive the relevant Smarr-like
relations for the mass and the tension which provide a connection
between the different characteristics of the solution. Let us
consider the expression for the tension (\ref{MT}). It is convenient
to reduce it to the factor space $\hat{M}$ by acting with the
Killing field $\eta = \frac{\partial}{\partial \phi}$ associated
with the azimuthal symmetry of the two-dimensional sphere at
infinity \cite{Yazad:2011}

\begin{eqnarray}
{\cal{T}}L =   {L\over 8} \int_{Arc(\infty)} \left[i_\eta i_k \star d\xi - 2i_\eta i_\xi \star d k \right]
\end{eqnarray}
The integration is now performed over the semicircle representing the boundary of the two-dimensional factor space at infinity.
Using the Stokes' theorem the integral can be further expanded into a bulk term over $\hat{M}$ and an integral over the rest of the boundary
of the factor space, which is represented by the interval structure $I_i$

\begin{eqnarray}
{\cal{T}}L = {L\over 8} \int_{\hat{M}} \left[d i_\eta i_k \star d\xi
- 2d i_\eta i_\xi \star d k \right] - {L\over 8}\sum_i \int_{I_i}
\left[i_\eta i_k \star d\xi - 2i_\eta i_\xi \star d k \right] .
\end{eqnarray}

If we take into account the definition of the intrinsic mass of the black hole (\ref{MH}) and the fact that the 1-form $i_\eta i_k \star d\xi$
vanishes along the left and right semi-infinite intervals $I_L$ and $I_R$, we obtain

\begin{eqnarray}\label{TL}
{\cal{T}}L =  \frac{1}{2}M_H + {L\over 4} \int_{I_L\bigcup I_R}
i_\eta i_\xi \star d k + {L\over 8} \int_{\hat{M}} \left[d i_\eta i_k \star d\xi
- 2d i_\eta i_\xi \star d k \right].
\end{eqnarray}

Let us consider the bulk integral and use the Ricci-identity $d\star d K=2\star R(K)$ which applies for any Killing field $K$

\begin{eqnarray}
 {L\over 8} \int_{\hat{M}} \left[ d i_\eta i_k \star d\xi - 2d i_\eta i_\xi \star d k \right] &=&  {L\over 8} \int_{\hat{M}} \left[i_\eta i_k d\star d\xi - 2i_\eta i_\xi d\star d k \right] = \nonumber \\
&& {L\over 4} \int_{\hat{M}} \left[i_\eta i_k \star R(\xi) -
2i_\eta i_\xi \star R(k) \right].
\end{eqnarray}
We can further show from the field equations that for any Killing field it is satisfied
\begin{eqnarray}
\star R(k) = - 2e^{-2a\varphi} \left( -{2\over 3}i_{k}F\wedge \star F + {1\over 3} F\wedge i_{k}\star F \right).
\end{eqnarray}

Applying this relation for the Killing fields $\xi$ and $k$ and considering the explicit form of the electromagnetic field we obtain

\begin{eqnarray}
i_\eta i_k \star R(\xi) - 2i_\eta i_\xi \star R(k) = -2\left( i_k F
\wedge i_\eta i_\xi e^{-2a\varphi}\star F + i_\eta F \wedge i_k
i_\xi e^{-2a\varphi}\star F\right).
\end{eqnarray}

Thus the bulk term becomes

\begin{eqnarray}
 &&{L\over 4} \int_{\hat{M}} \left[i_\eta i_k \star R(\xi) - 2i_\eta i_\xi \star R(k) \right] =
  - {L\over 2} \int_{\hat{M}}\left[ i_k F \wedge i_\eta i_\xi e^{-2a\varphi}\star F + i_\eta F \wedge i_k i_\xi e^{-2a\varphi}\star F\right] = \nonumber \\
 && - {L\over 2} \int_{\hat{M}}\left[ d\Phi_k \wedge i_\eta i_\xi e^{-2a\varphi}\star F + d\Phi_\eta \wedge i_k i_\xi e^{-2a\varphi}\star F\right].
\end{eqnarray}

We can further simplify the expression using Stokes' theorem and considering that the 1-form $i_ki_\xi \star F$ tends to zero at infinity,
as well as that the integral over the horizon vanishes

\begin{eqnarray}
 &&- {L\over 2} \int_{\hat{M}}\left[ d\Phi_k \wedge i_\eta i_\xi e^{-2a\varphi}\star F + d\Phi_\eta \wedge i_k i_\xi e^{-2a\varphi}\star F\right]=
  - {L\over 2} \int_{Arc(\infty)} \Phi_k i_\eta i_\xi e^{-2a\varphi}\star F - \nonumber \\
 && - {L\over 2} \int_{I_L\bigcup I_R}\left[ \Phi_k i_\eta i_\xi e^{-2a\varphi}\star F + \Phi_\eta i_k i_\xi e^{-2a\varphi}\star F\right].
\end{eqnarray}

Substituting this expressions in equation (\ref{TL}) we obtain

\begin{eqnarray}
{\cal{T}}L &=&  \frac{1}{2}M_H + {L\over 4} \int_{I_L\bigcup I_R}i_\eta i_\xi \star d k
 - {L\over 2} \int_{Arc(\infty)} \Phi_k  i_\eta i_\xi e^{-2a\varphi}\star F - \nonumber \\
&&- {L\over 2} \int_{I_L\bigcup I_R}\left[ \Phi_k  i_\eta i_\xi
e^{-2a\varphi}\star F + \Phi_\eta i_k i_\xi e^{-2a\varphi}\star
F\right],
\end{eqnarray}
which can be also represented as

\begin{eqnarray}
{\cal{T}}L &=&  \frac{1}{2}M_H + {L\over 4}r_\infty \int_{I_L}d\chi - {L\over 4}r_\infty \int_{I_R}d\chi + \nonumber \\
&+&{L\over 2} \int_{I_L}\left(\Phi_\eta + r_\infty \Phi_k \right)d{\cal{B}} + {L\over 2} \int_{I_R}\left(\Phi_\eta - r_\infty \Phi_k \right)d{\cal{B}} - \nonumber \\
&-& {L\over 2} \int_{Arc(\infty)} \Phi_k i_\eta i_\xi e^{-2a\phi}\star F .
\end{eqnarray}

From the definition (\ref{Nut potential}) of the nut potential it
follows that the nut potential is constant on the horizon, provided
the horizon is bifurcational, and we will denote its value by
$\chi$. Using the definition of the nut charge (\ref{Nut}) and the
fact  that the nut potential vanishes at infinity the last relation
is reduced to

\begin{eqnarray}
{\cal{T}}L &=&  \frac{1}{2}M_H + L N \chi + \nonumber \\
&+&{L\over 2} \int_{I_L}\left(\Phi_\eta + r_\infty \Phi_k \right)d{\cal{B}}
 + {L\over 2} \int_{I_R}\left(\Phi_\eta - r_\infty \Phi_k \right)d{\cal{B}} - \nonumber \\
&-& {L\over 2} \int_{Arc(\infty)} \Phi_k i_\eta i_\xi
e^{-2a\varphi}\star F .
\end{eqnarray}

The explicit form (\ref{solution}) of the electromagnetic potentials
$\Phi_k$ and $\Phi_\eta$  and the alignment of the left and right
semi-infinite intervals imply that $\Phi_\eta + r_\infty \Phi_k = 0$
on $I_L$, and $\Phi_\eta - r_\infty \Phi_k = 0$ on $I_R$. Since
$d{\cal B}$ is is regular on the factor space $\hat{M}$, the
relevant integrals vanish. It remains to calculate the integral
over the semicircle at infinity. We have

\begin{eqnarray}
{L\over 2} \int_{Arc(\infty)}\!\!\! \Phi_k i_\eta i_\xi
e^{-2a\varphi}\star F ={L\over
2}\Phi_{k}(\infty)\int_{Arc(\infty)}\!\!\!
 i_\eta i_\xi e^{-2a\varphi}\star F= \frac{1}{2}\Psi \!\int_{Arc(\infty)}
 i_\eta i_\xi e^{-2a\varphi}\star F
\end{eqnarray}
where $\Psi$ is the magnetic flux defined in (\ref{flux}).

In analogy with magnetostatics  it is natural to interpret the
integral

\begin{eqnarray}
J =-\frac{1}{2}\int_{Arc(\infty)}e^{-2a\varphi}i_\eta i_\xi\star F=
\frac{1}{4\pi}\int_{S^2_\infty}e^{-2a\varphi}i_\xi\star F
\end{eqnarray}
as the effective current that serves as a source of the magnetic field.

Thus we obtain the Smarr-like relation for the tension in its final form

\begin{eqnarray}
{\cal{T}}L &=&  \frac{1}{2}M_H + L N \chi + \Psi J.
\end{eqnarray}
The effective current $J$ can be expressed via the potential
$\Gamma$ which is defined by $d\Gamma=e^{-2a\varphi} i_\xi
i_\eta\star F$ and is given explicitly by

\begin{eqnarray}
\Gamma = -\frac{\lambda}{2}\frac{r_0\cos\theta (r - r_+)}{(1 + \lambda^2)(r + r_0)},
\end{eqnarray}
where it is normalized appropriately in order to vanishes on the horizon.

It is easy to see that the  effective current $J$ is connected to the restriction of the potential $\Gamma$ to the boundary of the factor space at infinity $Arc(\infty)$ as
\begin{eqnarray}
J=\frac{1}{2} \left[ \Gamma(\theta=\pi)\mid_{Arc(\infty)} - \Gamma(\theta=0)\mid_{Arc(\infty)}\right].
\end{eqnarray}

In a similar way if we take advantage of the Komar integral definition of the ADM mass we can derive the Smarr-like relation for the mass

\begin{eqnarray}
M &=&  M_H + \frac{1}{2}L N \chi .
\end{eqnarray}

\section*{Acknowledgements}
S.Y.  would like to thank the Alexander von Humboldt Foundation for
the support, and the Institut f\"ur Theoretische Astrophysik
T\"ubingen for its kind hospitality. The partial support by the
Bulgarian National Science Fund under Grants DO 02-257 and DMU-03/6
is also gratefully  acknowledged.

\end{document}